\pdfoutput=1
\documentclass[twocolumn,amsmath,amssymb,aps,secnumarabic,%
    nofootinbib,superscriptaddress,letterpaper]{revtex4-1}
\usepackage{mathptmx,amsthm,tikz,subfigure}
\usetikzlibrary{decorations.markings}
\usetikzlibrary{arrows}

\newtheorem{theorem}{Theorem}[section]

\newtheorem{corollary}[theorem]{Corollary}
\theoremstyle{remark}
\newtheorem*{remark}{Remark}

\newcommand{\C}{\mathbb{C}}
\renewcommand{\tensor}{\otimes}

\newcommand{\cH}{\mathcal{H}}
\newcommand{\ket}[1]{|#1\rangle}

\newcommand{\braket}[1]{\langle#1\rangle}

\newcommand{\vF}{\vec{F}}

\newcommand{\eps}{\epsilon}

\newcommand{\BPP}{\mathsf{BPP}}
\newcommand{\BQP}{\mathsf{BQP}}
\newcommand{\FH}{\mathsf{FH}}
\newcommand{\QNC}{\mathsf{QNC}}
\newcommand{\SQCL}{\mathsf{SQCL}}
\newcommand{\AQCL}{\mathsf{AQCL}}

\newcommand{\thm}[1]{Theorem~\ref{#1}}
\renewcommand{\sec}[1]{Section~\ref{#1}}

\newcommand{\cor}[1]{Corollary~\ref{#1}}

\newcommand{\eq}[2]{\begin{equation}\label{#1}#2\end{equation}}

\newenvironment{fullfigure}[2]
    {\begin{figure}[htb]\def\ffa{#1}\def\ffb{#2}}
    {\caption{\ffb.}\label{\ffa}\end{figure}}
\newenvironment{fullfigure*}[2]
    {\begin{figure*}[tb]\def\ffa{#1}\def\ffb{#2}}
    {\caption{\ffb.}\label{\ffa}\end{figure*}}
\newcommand{\fig}[1]{Figure~\ref{#1}}

\colorlet{darkred}{red!70!black}
\colorlet{darkgreen}{green!70!black}
\colorlet{darkbrown}{brown!70!black}
\colorlet{lightblue}{blue!70!white}
\colorlet{lightgray}{white!70!black}
\colorlet{lightred}{red!60!lightgray}

\begin{document}
\title{Contagious error sources would need time travel to prevent
    quantum computation}

\author{Gil Kalai}
\email{kalai@math.huji.ac.il}
\thanks{Supported by ERC advanced grant 320924, ISF grant 768/12,
    and NSF grant DMS-1300120}
\affiliation{Hebrew University in Jerusalem, Israel}
\author{Greg Kuperberg}
\email{greg@math.ucdavis.edu}
\thanks{Partly supported by NSF grant CCF-1319245}
\affiliation{University of California, Davis}

\begin{abstract}
We consider an error model for quantum computing that consists of
``contagious quantum germs" that can infect every output qubit when at
least one input qubit is infected.  Once a germ actively causes error,
it continues to cause error indefinitely for every qubit it infects, with
arbitrary quantum entanglement and correlation.  Although this error model
looks much worse than quasi-independent error, we show that it reduces
to quasi-independent error with the technique of quantum teleportation.
The construction, which was previously described by Knill, is that every
quantum circuit can be converted to a mixed circuit with bounded quantum
depth.   We also consider the restriction of bounded quantum depth from
the point of view of quantum complexity classes.
\end{abstract}
\maketitle

\section{Introduction}

Is quantum computation realistic even in principle?   If we accept quantum
mechanics (more precisely, quantum probability), then at the theoretical
level this question is usually interpreted as the fault tolerance problem:
Can a quantum computer still work if all of its gates and qubits are noisy?
There are by now various fault-tolerance theorems for quantum computation,
which establish that reliable quantum computation is indeed possible in
principle assuming that the noise present in different qubits or gates is
quasi-independent \cite{KLZ:resilient,AB:constant,Kitaev:imperfect}, and
is below some threshold error rate.  This threshold is called the fault
tolerance constant.  Thus, any remaining doubt that quantum computation
is possible in principle reduces to one of three possibilities:
\begin{description}
\item[1.] Quantum probability is not exactly true.
\item[2.] The fault tolerance constant is unattainable.
\item[3.] The quasi-independence assumption is too optimistic.
\end{description}

In this note, we will consider noise models with a relaxed version of the
quasi-independence assumption, namely \emph{contagious noise}.  It seems
possible that each qubit in a quantum computer might not just be noisy,
but carry with it a noise source, a contagious ``bug", that spreads to
all of the output qubits of each quantum gate.   Each bug could get worse
over time.  Worse still, the descendants of the bug could be correlated
and thus violate the quasi-independence assumption.  If a quantum gate has
two different bugs among its inputs, the bugs might also interact and make
new bugs.  Such possibilities come to mind given that one of the first
bugs in the history of modern computing was an actual bug, a small moth
\cite{Hopper:bug}.  That bug was no longer interacting with anything other
than the relay switch where it had died.   A ``bug" can also mean a germ;
at least in biological computers, germs can both replicate and affect data.

More realistically, contagious error is related to some forms of leakage
error, where what was the state of a qubit leaves the qubit Hilbert space
and enters a larger Hilbert space.  Knill has noted that leakage
error is implicitly solved by teleportation \cite{Knill:scalable}, which
is also the method that we will use. Leakage error is generally thought of
as a measured error; if it is measured, it amount to a qubit erasure and
the qubit can be reset.  However, before it is measured, leakage error
can be contagious, since the effect of a quantum gate is undefined for
leaked states.

In this article, we propose a mutual generalization of contagious germs and 
leakage error which we call \emph{contagious quantum germs}.  If a qubit has 
a Hilbert space $\cH_Q \cong \C^2$, then we attach to it another Hilbert 
space $\cH_G$, the germ state space, so that its total state space is $\cH_Q 
\tensor \cH_G$.  At each time step, each qubit interacts with its germ and 
its germ evolves.  At each gate, all of the germs of the input qubits 
interact in some way to make the germs of the output qubits.  The only 
restrictions are that each fresh qubit is created with an independent germ, 
that the effect of a germ on its qubit is bounded above for the first few 
steps of its life, and that classical bits do not carry germs.  Note that 
quantum germs not only strictly generalize leakage error, but can also do 
things that would be peculiar for leakage error. For instance, once 
activated, they can spread insidiously with no possibility of being 
detected, and then hit hard everywhere that they have spread.   We have no 
argument that all types of quantum germs are realistic, only that they 
include various possibly realistic noise models as special cases.

\begin{theorem} With a constant overhead factor, every quantum circuit
can be re-encoded so that noise from contagious quantum germs becomes
quasi-independent.
\label{th:main} \end{theorem}

As a corollary, we can then apply the standard fault-tolerance theorem
to conclude that quantum computation is possible in our model with
polylogarithmic overhead.

A more detailed version of \thm{th:main} is stated as \thm{th:rig} and
\cor{c:replace}, using the formalism defined in \sec{s:rig}.

A closely related result is the ``one-way" quantum fault tolerance proposal 
of Raussendorf, Harrington, and Goyal \cite{RHG:oneway,RHG:cluster}. This is 
a specific encoded computation with a high fault-tolerance constant, and 
which is one-way in the sense that after an initial encoded state is 
created, all of the computation is carried out by (adaptive) one-qubit 
measurements.   Since in addition all of the parity checks of their code 
have bounded weight, their circuits automatically have bounded depth.  Our 
\thm{th:main} is both weaker and more general than the RHG construction.  
Without obtaining any new bound on the fault tolerance constant, it implies 
that any method of quantum fault tolerance whatsoever can be re-encoded in 
bounded quantum depth.

The idea of our proof of \thm{th:main} is not original.  The basic
construction uses quantum teleportation \cite{BBCJPW:teleport}; it might
first have been published by Knill \cite{Knill:postsel}.  We state the
construction in terms of mixed quantum-classical circuits.

\begin{theorem}[Knill] With a constant overhead factor, every quantum
circuit (or mixed classical-quantum circuit) can be re-encoded as a mixed
classical-quantum circuit with bounded quantum depth.
\label{th:knill} \end{theorem}

The surprising property of a mixed circuit with quantum teleportation is 
that, if we orient all of the qubit edges forward in time, then the qubit 
subgraph only needs to be weakly connected from its input to its output in 
order to transmit quantum information.   (Recall that directed graph is \emph
{strongly} connected from $a$ to $b$ if it has a directed path from $a$ to 
$b$; and \emph{weakly} connected if it merely has some connecting path with 
no restriction on the orientations of the edges.)  In a sense, quantum 
information can travel backward in time, as long as classical information 
transports the result forward in time.  Indeed, quantum teleportation is a 
reasonable description of \emph{any} mixed circuit that transmits quantum 
information via a weakly connected quantum subgraph; in this sense, quantum 
information that travels backwards in time is just a slogan for 
teleportation.  Theorems~\ref{th:main} and \ref{th:knill} then say that if 
we imagine that qubits (but not bits) are infected with contagious germs, 
then the germs would have to travel backward in time to prevent 
fault-tolerant quantum computation.

Given that we can loosely interpret a teleportation circuit as allowing 
quantum information to travel backwards in time, the question arises whether 
quantum germs could travel backwards in time in the same sense of 
teleportation.   The answer is no, unless either:
\begin{description}
\item[1.] Germs can infect classical bits.
\item[2.] Germs can travel on paths that are not in the circuit.
\end{description}
Given the many different ways that classical bits are encoded and 
communicated in practice, the first possibility seems implausible or at 
least avoidable.  The second possibility certainly can happens in the sense 
that noise may not be limited to a quantum circuit in example physical 
implementations.  In this case, the noise does not need to travel backwards 
in time either.  However, if correlated noise can spread anywhere with no 
causal restriction other than that it travels forward in time, then it is 
well-known that either classical or quantum fault tolerance is impossible.

The authors were led to consider the constructions considered here by
alternative error models proposed by the first author in which errors
are convolved, or smoothed, in time \cite{Kalai:adversarial}.  Our basic
observation led the first author to change his model to require convolution
both forward and backward in time \cite{Kalai:fail}.  This leads to issues
regarding causality that we will not discuss here.

\begin{remark}  If a mixed circuit is not even weakly quantumly connected,
then it is equivalent to a model in quantum information theory known as
``local operations and classical communication" \cite[\S12.5]{NC:qcqi}.
In particular, it is immediate that LOCC is weaker than full quantum
communication, since it leaves no way to create quantum entanglement
between weakly connected components of the circuit, and therefore no way
to violate Bell-type inequalities.
\end{remark}

To understand our hypothesis and our conclusion, it is important to
distinguish between sources of error and erroneous qubits (or bits).
An erroneous qubit is one whose state is different from what is intended in
a quantum algorithm.  If the intended state is a pure state $\ket{\psi}$
(or a density operator $\rho$), then the actual state might be some other
state $\ket{\psi'}$ (or a density operator $\rho'$).   Erroneous states
propagate through gates and through quantum teleportation.  In computer
science in general, this is called \emph{error propagation} and it is
why computers need classical or quantum error correction.  In fact, an
error can propagate through a mixed path of classical and quantum edges
in a mixed circuit; in particular, an error can propagate through the
classical bits used in quantum teleportation.  However, while errors
can be corrected, by our rules error-causing germs cannot be removed.
(Or, they can only be removed indirectly with teleportation.)

We will prove \thm{th:knill} in \sec{s:knill} and \thm{th:main} in
\sec{s:main}.  Finally, in \sec{s:complexity}, we give a complexity theory
interpretation of \thm{th:knill}.  One way to limit the power of quantum
computation is to only allow bounded-time layers of it in between classical
computation layers.  (We do not mean pseudo-classical operators that are
quantum but in the computational basis.  Rather we mean classical data
processing revealed to the environment.)  We remark that if this is done
asynchronously, then \thm{th:knill} implies that the resulting polynomial
complexity class is exactly $\BQP$.

\acknowledgments 

The authors would like to thank Dorit Aharonov, Andrew Childs, Joe 
Fitzsimons, Eleanor Rieffel, and especially Daniel Gottesman for useful 
conversations and corrections.

\section{Rigorous definitions}
\label{s:rig}

\subsection{Mixed circuits}
\label{s:mixed}

We consider the circuit model of computation.  As usual, a circuit is a kind
of acyclic, directed graph with labelled vertices which are called gates.
In defining classical circuits carefully enough to generalize them to quantum
circuits, we have to count bit copying as a gate with 1 input and 2 outputs.
Also, every type of circuit that we consider in this paper can be assumed
to be in a uniform circuit family, created by a classical Turing machine or
similar.

The circuits of interest to us have two kinds of circuit edges, classical
or bit edges, and quantum or qubit edges.   In order to understand what
a general mixed gate can do with a combination of bit and qubit inputs
and outputs, we can consider the \emph{hybrid quantum memory} model
\cite{Kuperberg:memory}.  In practice, we can simplify the definition of
a mixed circuit to the following types of gates:
\begin{description}
\item[1.] Deterministic classical gates acting on bits.
\item[2.] A measurement gate that converts a qubit to a bit.
\item[3.] Unitary quantum gates that may have classical control bits.
\item[4.] A gate that creates a fresh qubit in the state $\ket{0}$.
\end{description}
Since our circuits are uniformly generated, and for other reasons, we also
want a finite set of quantum gates that densely generate unitary groups
acting on qubits, such as the Hadamard and Toffoli gates.  However, rather
than proving something for every gate set, we interpret \thm{th:main}
as saying that there exists a set of gates such that the result holds
with a constant overhead factor for those gates.  (Changing gate sets
requires the Solovay-Kitaev theorem, which has polylogarithmic overhead;
we do not know that the stringent constant overhead factor can be satisfied
by every universal gate set.)  The only standard gates that we need for
quantum teleportation (\fig{f:teleport}), which is the only idea we need
to prove \thm{th:knill}, are Hadamard and CNOT gates, and 1-qubit Pauli
gates with classical control bits.

\fig{f:teleport} has an example of a mixed circuit, with the qubit edges
in red and the bit edges in blue.   In general a circuit has a \emph{total
depth}, which is the length of its longest directed path; and a \emph{quantum
depth}, which is the length of its longest directed path following only
qubit edges.  The graph of a mixed circuit has a \emph{quantum subgraph}
consisting only of its qubit edges.  (The specific gates used 
in the circuit are defined in \sec{s:knill}.)

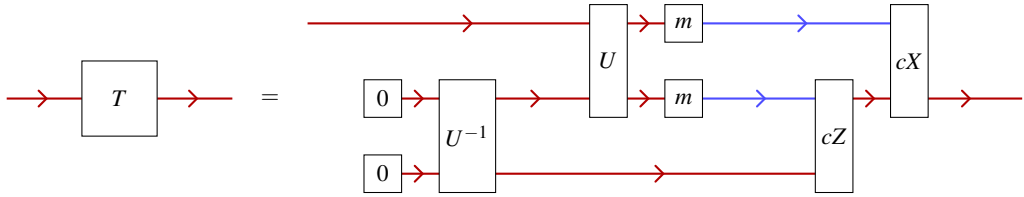
\begin{fullfigure*}{f:teleport}{The teleportation circuit $T$, with
    qubit edges in red and bit edges in blue.   The output is connected
    to the input by a path of qubit edges, but not by an \emph{oriented}
    path}
\begin{tikzpicture}[decoration={markings,
    mark=at position 0.55 with {\arrow{angle 90}}}]
\begin{scope}[thick]
\draw[darkred,postaction={decorate}] (-5,1) -- (-4,1);
\draw[darkred,postaction={decorate}] (-3,1) -- (-2,1);

\draw[darkred,postaction={decorate}] (0,0) -- (1,0);
\draw[darkred,postaction={decorate}] (1,0) -- (6,0);

\draw[darkred,postaction={decorate}] (0,1) -- (1,1);
\draw[darkred,postaction={decorate}] (1,1) -- (3,1);
\draw[darkred,postaction={decorate}] (3,1) -- (4,1);
\draw[lightblue,postaction={decorate}] (4,1) -- (6,1);
\draw[darkred,postaction={decorate}] (6,1) -- (7,1);
\draw[darkred,postaction={decorate}] (7,1) -- (8.5,1);

\draw[darkred,postaction={decorate}] (-1,2) -- (3,2);
\draw[darkred,postaction={decorate}] (3,2) -- (4,2);
\draw[lightblue,postaction={decorate}] (4,2) -- (7,2);
\end{scope}
\draw[fill=white] (-4,.5) rectangle (-3,1.5); \draw (-3.5,1) node {$T$};
\draw (-1.5,1) node {$=$};
\draw[fill=white] (-.25,-.25) rectangle (.25,.25); \draw (0,0) node {$0$};
\draw[fill=white] (-.25,.75) rectangle (.25,1.25); \draw (0,1) node {$0$};
\draw[fill=white] (.75,-.25) rectangle(1.5,1.25); \draw (1.125,.5) node {$U^{-1}$};
\draw[fill=white] (2.75,.75) rectangle(3.25,2.25); \draw (3,1.5) node {$U$};
\draw[fill=white] (3.75,.75) rectangle (4.25,1.25); \draw (4,1) node {$m$};
\draw[fill=white] (3.75,1.75) rectangle (4.25,2.25); \draw (4,2) node {$m$};
\draw[fill=white] (5.75,-.25) rectangle(6.25,1.25); \draw (6,.5) node {$cZ$};
\draw[fill=white] (6.75,.75) rectangle(7.25,2.25); \draw (7,1.5) node {$cX$};
\end{tikzpicture}
\end{fullfigure*}

\subsection{Quantum germs}
\label{s:germs}

As usual $U(n)$ is the group of unitary $n \times n$ matrices; let $M(n)$
be the vector space of all $n \times n$ matrices.  If $\cH$ is a Hilbert
space, then we let $U(\cH)$ be the corresponding abstract unitary group,
and we let $M(\cH)$ be the abstract space of all operators on $\cH$.
The algebra $M(\cH)$ comes with an operator norm; by definition
\[ ||A|| = \sup_{\braket{\psi|\psi} = 1} \braket{\psi|A|\psi}. \]
If $\cH$ is infinite-dimensional, then technically we take $M(\cH)$
to be the bounded operators, meaning those with finite operator norm.
Recall in this case that a state $\rho$ on $\cH$ is a positive semi-definite
trace-class operator.  (More precisely, if $\rho$ is positive semi-definite,
then it is \emph{trace class} if the trace of its matrix defined using
any orthonormal basis of $\cH$ is a convergent series.)

As mentioned in the introduction, each qubit has a Hilbert space $\cH_Q
\cong \C^2$, and a separate germ Hilbert space $\cH_G$ that could even
be infinite-dimensional.   If $C$ is a quantum circuit (or a mixed
circuit), we expand it to an infected circuit $C'$ as follows:  If 
$C$ creates a qubit with the state $\ket{0} \in \cH_Q$, then $C'$
also creates a germ in some initial state $\ket{g_0} \in \cH_G$.
For each gate
$$G \in U(\cH_Q^{(1)} \tensor \cH_Q^{(2)} \tensor \cdots \tensor \cH_Q^{(k)})$$
that arises in $C$, there is a corresponding germ-mixing operator
$$M \in U(\cH_G^{(1)} \tensor \cH_G^{(2)} \tensor \cdots \tensor \cH_G^{(k)})$$
that mixes the germ states.  Finally, each edge of the circuit $C$ is
replaced with an error operator
$$E \in U(\cH_G \tensor \cH_Q).$$
Also, we do not assume that the operators $M$ and $E$ and the states
$\ket{g_0}$ are the same at different positions in $C$.   They must satisfy
the error bound \eqref{e:bound} below, but otherwise they can be different
each time and they can be chosen adversarially rather than randomly.

The operators $E$ are subject to an error bound which we explain carefully.
Recall the relation
$$M(\cH_G \tensor \cH_Q) \cong M(\cH_G) \tensor M(\cH_Q).$$
Recall that $M(\cH_Q) \cong M(2)$ can be given a Pauli basis
\begin{align*}
P_0 &= I = \begin{pmatrix} 1 & 0 \\ 0 & 1 \end{pmatrix}, &
P_1 &= X = \begin{pmatrix} 0 & 1 \\ 1 & 0 \end{pmatrix} \\
P_2 &= Y = \begin{pmatrix} 0 & -i \\ i & 0 \end{pmatrix}, &
P_3 &= Z = \begin{pmatrix} 1 & 0 \\ 0 & -1 \end{pmatrix}
\end{align*}
using any isomorphism $\cH_Q \cong \C^2$.  Then we can think of an
operator $E \in M(\cH_G \tensor \cH_Q)$ as a superposition of operators
acting only on $\cH_G$:
\eq{e:superpos}{E = \sum_{j=0}^3 E_j \tensor P_j.}
The fact that $E$ is unitary implies that $||E_j|| \le 1$, which we can
read as saying that each $E_j$ is subunitary; this will be useful in
the proof of \thm{th:main}.

As a warm-up to the main argument, suppose that the germ at a given qubit
edge has a pure state $\ket{g} \in \cH_G$.  Then the partial evaluation
of $E$ on $\ket{g}$ gives us a vector
$$\vF \in \cH_G \tensor M(\cH_Q),$$
which then decomposes as
\eq{e:klz}{\vF = \sum_{j=0}^3 \ket{f_j} \tensor P_j.}
Here each $\ket{f_j}$ is a non-normalized state representing a vector-valued
amplitude of the error mode.  Following Knill, Laflamme, and Zurek \cite[\S
I.B]{KLZ:resilient}, we can define the size of this error as the sum of
the norms of the output germ states $\ket{f_j}$ other than the term for
the identity.  In other words, we can define an error seminorm
$$||\vF|| = \sum_{j=1}^3 \sqrt{\braket{f_j|f_j}}.$$
One subtle but standard point, which will be relevant in all of our error
bounds, is that the vectors such as $\ket{f_j}$ need not be orthogonal.
If they are orthogonal, then the different errors to which they are attached
are stochastic; if they are parallel, then the errors are coherent or
``stoquastic".

If $||\vF||$ is large, it means that the error operator $E$ has a large
effect on its qubit $Q$.   We would like to bound $||\vF||$; however for
two reasons, we will not do this in all cases.   The first reason is that
the state $\rho_G$ in general comes from a pure state that is entangled
between many germs and computational qubits as well.  The second reason
is that we assume that if a germ creates an error in a qubit, then it is
activated and can cause later errors with high probability.

We address the first issue, and clarify the second one, by passing to a
multilinear expansion of all of the error operators using \eqref{e:klz}.
Instead of directly considering the full vector state of all of the germs and
the errors they cause, we can instead consider the amplitude contribution of
any particular pattern of Pauli errors.  The total error is a superposition
of all of these patterns.  To prove Theorem~\ref{th:main}, we will bound each
term of the superposition separately, and then sum to get the total bound.

If there are $N$ edges, then
we expand all possible errors across the circuit $C$:
\eq{e:mpauli}{ \vF = \sum_{J=0}^{4^N-1} \ket{f_J} \tensor P_J. }
Here $P_J$ is a \emph{multi-Pauli} operator, a tensor product of Pauli
operators including the identity.  We consider the partial ordering on
qubit edges in which $q_1 \prec q_2$ if there is a directed path from $q_1$
to $q_2$.  Then with respect to this partial ordering, some of the Pauli
factors of $P_J$ are the earliest among those that are not the identity.
We call these qubit edges \emph{locally first diseased} (in superposition).

If $q$ is locally first diseased, then all of the germs that ever interacted
with the one at $q$ have an entangled state
$$\ket{g} \in \cH_G^{(1)} \tensor \cH_G^{(2)}
    \tensor \cdots \tensor \cH^{(k)}.$$
The state $\ket{g}$ is formed from various initial states $\ket{g_0}$
with germ-mixing operators $M$ and error components $E_j$ acting on
them.  Since each $M$ is unitary and each $E_j$ is subunitary, we learn
that $|\braket{g|g}| \le 1$.  We can make an error vector $\vF$
\eqref{e:klz} in the same way as in the warmup case, except using a state
$\ket{g}$ of many germs rather than one germ.   We assume an upper bound
\eq{e:bound}{||\vF|| < \eps(n),}
if the quantum depth of the part of the circuit $C$ that leads to $q$ is at
most $n$.  We assume that $\eps(n)$ is a small number when $n$ is small.
Otherwise, if the best upper bound $\eps(n)$ is large for small $n$,
then even small quantum circuits are unreliable; with enough such noise,
there is no clear reason to expect quantum computation to be possible.

\section{Proof of \thm{th:knill}}
\label{s:knill}

\begin{proof} The theorem reduces to the existence of quantum teleportation.
Quantum teleportation is a mixed quantum-classical circuit $T$ that has
one qubit input and one qubit output, and no quantum path from the input to
the output.  Moreover, the circuit computes the identity: The output agrees
with the input and it even inherits any entanglement that the input had
with other qubits.  The teleportation circuit is given in \fig{f:teleport};
a simplified version is given in \fig{f:telesimp}.

As mentioned, our convention for all diagrams is that qubit edges are
red and bit edges are blue.  The gates used in the expanded circuit in
\fig{f:teleport} are as follows:
\begin{itemize}
\item[1.] The gate $0$ creates a qubit in the state $\ket{0}$.
\item[2.] The gate $m$ measures a qubit in the computational basis
and outputs a bit.  The qubit input is destroyed.
\item[3.] The gate $U$ is unitary with the following action:
\begin{align*} U\ket{00} &= \frac{\ket{00}+\ket{01}}{\sqrt{2}} &
U\ket{01} &= \frac{\ket{10}-\ket{11}}{\sqrt{2}} \\
U\ket{10} &= \frac{\ket{10}+\ket{11}}{\sqrt{2}} &
U\ket{11} &= \frac{\ket{00}-\ket{01}}{\sqrt{2}}.
\end{align*}
It can be created as a CNOT gate
\[ \ket{x,y} \mapsto \ket{x+y,y} \]
followed by a Hadamard gate 
\[ H = \frac1{\sqrt{2}} \begin{pmatrix} 1 & 1 \\ 1 & -1 \end{pmatrix} \]
applied to the second qubit.
\item[4.] The gate $cX$ applies the one-qubit operator $X$ if its bit
input is 1, and the identity $I$ if it is 0.
\item[5.] The gate $cZ$ applies the one-qubit operator $Z$ if its bit
input is 1, and the identity $I$ if it is 0.
\end{itemize}
If the circuit $T$ is inserted at every edge of a circuit $C$ to make $C'$,
then the longest directed qubit path in $C'$ is 6 edges.  An initialized
qubit lasts for 4 edges inside $T$.  The output of $T$ can be the input
to some gate $G$ in $C$, and then the output of $G$ can become the input
to another copy of $T$ and last for 2 more edges to make 6 total.
\end{proof}

It is a celebrated result that the teleportation circuit $T$ computes
the identity, even when its input is entangled.  In order to prove this,
it suffices to check that it is the identity for any spanning set of
density operators.  For example if the circuit preserves the vector states
$\ket{0}$, $\ket{1}$, $\ket{\pm}$, and $\ket{\pm i}$, then it also preserves
all six corresponding density operators, which implies that it preserves
all density operators.  It is easy to check that these states are indeed
preserved by $T$.

\section{Proof of \thm{th:main}}
\label{s:main}

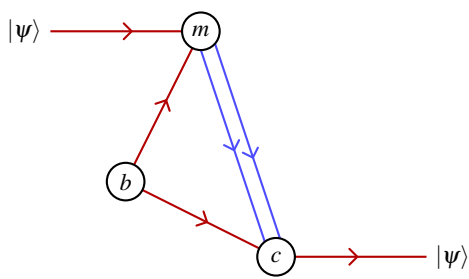
\begin{fullfigure}{f:telesimp}{A simplified diagram of the teleportation
    circuit in \fig{f:teleport}, in which the gates are combined as much
    as possible.  The gate $b$ creates a Bell pair; the gate $m$ measures
    two qubits in an entangled basis, and the gate $c$ is a unary quantum
    gate controlled by two bits}
\begin{tikzpicture}[thick,decoration={markings,
    mark=at position 0.55 with {\arrow{angle 90}}}]
\draw[darkred,postaction={decorate}] (-1,3) -- (1,3);
\draw[darkred,postaction={decorate}] (0,1) -- (1,3);
\draw[darkred,postaction={decorate}] (0,1) -- (2,0);
\draw[lightblue,postaction={decorate}] (.9,3.05) -- (1.9,0.05);
\draw[lightblue,postaction={decorate}] (1.15,2.95) -- (2.15,-0.05);
\draw[darkred,postaction={decorate}] (2,0) -- (4,0);
\draw[fill=white] (0,1) circle (.25); \draw (0,1) node {$b$};
\draw[fill=white] (1,3) circle (.25); \draw (1,3) node {$m$};
\draw[fill=white] (2,0) circle (.25); \draw (2,0) node {$c$};
\draw[anchor=east] (-1,3) node {$\ket{\psi}$};
\draw[anchor=west] (4,0) node {$\ket{\psi}$};
\end{tikzpicture}
\end{fullfigure}

We will actually prove two different results that both fit the words of
\thm{th:main}.  

\begin{theorem} Let $C$ be a circuit with bounded quantum depth $n$ and gates
that act on at most $k$ qubits.   If $C$ is infected with quantum germs with
error bound $\delta(n)$ at depth at most $n$, then $C$ has quasi-independent
error with bound $\eps$ depending on $n$, $k$, and $\delta(n)$.  Moreover,
$\eps \to 0$ as $\delta(n) \to 0$.
\label{th:rig} \end{theorem}

\begin{corollary} If $C$ is a circuit with unbounded quantum depth, then it
can be replaced by an equivalent circuit $C'$, so that if $C'$ is infected
with quantum germs, then the result is equivalent to quasi-independent
noise in $C$.
\label{c:replace} \end{corollary}
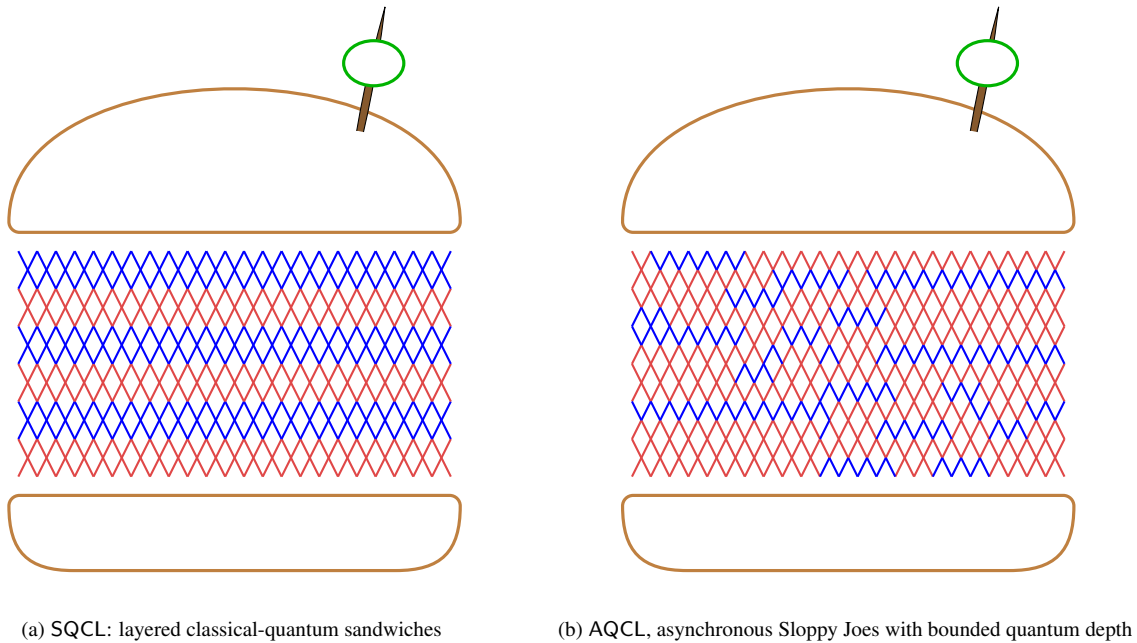
\begin{fullfigure*}{f:sandwich}{A schematic comparison of two
 kinds of mixed quantum-classical circuits}
\subfigure[\;$\SQCL$: layered classical-quantum sandwiches]{
\begin{tikzpicture}
\useasboundingbox (-4,-1.5) rectangle (4,7);
\draw[rounded corners,brown,very thick]
    (0,3.5) -- (3,3.5) .. controls (3,6) and (-3,6) .. (-3,3.5) -- (0,3.5);
\draw[rounded corners,brown,very thick] (0,0) --
    (-3,0) .. controls (-3,-.75) and (-2.75,-1) .. (-2,-1) --
    (2,-1) .. controls (2.75,-1) and (3,-.75) .. (3,0) -- (0,0);
\draw[fill=darkbrown] (1.5,4) to[bend right=4] (2,6.5) to[bend right=4] (1.5,4);
\draw[fill=white,draw=darkgreen,very thick] (1.85,5.75) ellipse (.4 and .3);
\draw[fill=white,draw=none] (1.6,4.4) circle (.45);
\foreach \x in {-2.75,-2.5,...,2.75}
    \foreach \y/\c in {.25/lightred,.75/blue,1.25/lightred,1.75/blue,
        2.25/lightred,2.75/blue} {
        \draw[\c,thick] (\x,\y) +(-.125,0) -- +(.125,.5);
        \draw[\c,thick] (\x,\y) +(.125,0) -- +(-.125,.5); }
\end{tikzpicture}}
\subfigure[\;$\AQCL$, asynchronous Sloppy Joes with bounded quantum depth]{
\begin{tikzpicture}
\useasboundingbox (-4,-1.5) rectangle (4,7);
\draw[rounded corners,brown,very thick]
    (0,3.5) -- (3,3.5) .. controls (3,6) and (-3,6) .. (-3,3.5) -- (0,3.5);
\draw[rounded corners,brown,very thick] (0,0) --
    (-3,0) .. controls (-3,-.75) and (-2.75,-1) .. (-2,-1) --
    (2,-1) .. controls (2.75,-1) and (3,-.75) .. (3,0) -- (0,0);
\draw[fill=darkbrown] (1.5,4) to[bend right=4] (2,6.5) to[bend right=4] (1.5,4);
\draw[fill=white,draw=darkgreen,very thick] (1.85,5.75) ellipse (.4 and .3);
\draw[fill=white,draw=none] (1.6,4.4) circle (.45);
\begin{scope}[thick,xscale=.125,yscale=.25,shift={(-23,1)}]
\foreach \x/\y in {
    44/4, 45/9, 33/3, 36/8, 34/8, 39/11, 27/1, 29/11, 16/0, 9/9, 23/9,
    14/8, 12/8, 5/1, 40/4, 41/1, 35/3, 24/2, 25/3, 19/1, 8/0, 42/8, 43/5,
    45/7, 32/4, 33/1, 37/11, 38/0, 39/5, 27/3, 28/4, 31/9, 29/9, 16/2,
    20/8, 18/8, 22/2, 23/11, 11/1, 12/10, 13/11, 0/0, 2/10, 41/7, 42/4,
    43/9, 41/11, 32/8, 35/5, 37/7, 25/1, 30/0, 31/5, 8/2, 14/2, 3/1, 6/2,
    43/7, 44/0, 45/5, 33/7, 37/9, 39/7, 27/5, 31/11, 29/7, 16/4, 17/1,
    18/6, 21/11, 23/5, 10/10, 12/4, 0/2, 4/8, 40/8, 41/5, 42/2, 43/11,
    35/7, 7/9, 37/5, 24/6, 25/7, 27/9, 33/11, 31/7, 16/8, 19/5, 21/7, 8/4,
    9/1, 10/6, 14/0, 15/5, 4/4, 44/2, 33/5, 34/2, 38/4, 39/1, 27/7, 28/0,
    29/5, 16/6, 18/4, 21/9, 22/6, 23/7, 10/8, 12/6, 15/11, 13/7, 0/4, 1/1,
    4/10, 2/6, 6/0, 7/5, 42/0, 35/9, 33/9, 36/2, 38/8, 25/5, 27/11, 30/4,
    31/1, 17/11, 21/5, 8/6, 10/4, 15/7, 3/5, 4/6, 45/1, 6/10, 39/3, 29/3,
    17/5, 18/2, 23/1, 12/0, 0/6, 2/4, 5/9, 6/6, 41/9, 44/8, 35/11, 37/1,
    24/10, 25/11, 31/3, 19/9, 17/9, 21/3, 8/8, 9/5, 10/2, 14/4, 15/1,
    0/10, 1/11, 4/0, 5/5, 40/0, 29/1, 18/0, 23/3, 12/2, 1/5, 2/2, 6/4,
    7/1, 43/1, 45/11, 25/9, 39/9, 30/8, 28/8, 19/11, 21/1, 8/10, 10/0,
    14/10, 3/9, 1/9, 4/2, 6/8, 41/3, 35/1, 13/1, 2/0}
    \draw[lightred] (\x,\y) -- +(1,1);
\foreach \x/\y in {
    16/9, 19/4, 6/9, 20/7, 8/5, 9/0, 14/1, 15/4, 3/2, 4/5, 45/2, 34/3,
    40/11, 38/5, 39/0, 28/1, 29/4, 17/6, 18/5, 22/7, 23/6, 11/4, 12/7,
    13/6, 0/5, 1/0, 5/10, 6/1, 7/4, 42/1, 35/8, 33/8, 37/2, 24/9, 38/9,
    26/3, 30/5, 31/0, 16/11, 19/6, 20/1, 22/11, 9/6, 10/5, 11/8, 14/7,
    0/9, 3/4, 5/6, 45/0, 34/1, 28/3, 17/4, 22/5, 11/6, 12/1, 13/4, 1/6,
    2/5, 5/8, 7/6, 44/9, 24/11, 26/1, 30/11, 19/8, 17/8, 21/2, 8/9, 22/9,
    9/4, 11/10, 15/0, 0/11, 3/6, 1/10, 4/1, 5/4, 40/1, 29/0, 18/1, 23/2,
    13/2, 1/4, 6/5, 7/0, 43/0, 44/11, 32/1, 34/11, 39/8, 30/9, 28/9, 9/10,
    10/1, 14/11, 15/2, 5/2, 40/3, 24/1, 13/0, 2/1, 7/2, 42/11, 43/2, 44/5,
    45/8, 32/3, 36/9, 34/9, 38/3, 28/11, 16/1, 18/11, 41/8, 9/8, 3/10,
    5/0, 40/5, 41/0, 42/7, 35/2, 36/5, 24/3, 25/2, 30/3, 19/0, 8/1, 42/9,
    43/4, 44/7, 32/5, 36/11, 34/7, 38/1, 39/4, 26/11, 28/5, 31/8, 29/8,
    17/2, 20/9, 18/9, 22/3, 7/10, 23/10, 11/0, 12/11, 13/10, 0/1, 40/7,
    7/8, 42/5, 43/8, 32/9, 36/7, 24/5, 26/7, 30/1, 31/4, 19/2, 20/5, 9/2,
    3/0, 44/1, 45/4, 32/7, 34/5, 37/8, 38/7, 26/9, 28/7, 16/5, 17/0, 20/11,
    22/1, 11/2, 15/8, 13/8, 1/2, 4/9, 2/9, 40/9, 41/4, 32/11, 36/1, 37/4,
    24/7, 38/11, 25/6, 26/5, 27/8, 30/7}
    \draw[lightred] (\x,\y) +(0,1) -- +(1,0);
\foreach \x/\y in {
    7/3, 11/11, 17/7, 43/3, 24/4, 3/7, 26/6, 32/2, 36/6, 11/5, 42/10, 38/2,
    19/3, 20/4, 5/11, 28/10, 9/3, 40/10, 36/4, 18/10, 37/3, 24/8, 3/11,
    26/2, 32/6, 34/4, 16/10, 13/3, 19/7, 42/6, 20/0, 26/8, 22/10, 28/6,
    9/7, 26/10, 11/9, 14/6, 20/10, 44/10, 0/8, 22/0, 32/0, 5/7, 11/3, 34/10,
    15/9, 13/9, 1/3, 2/8, 34/0, 7/11, 28/2, 9/11, 32/10, 36/0, 30/6, 15/3,
    38/10, 44/6, 22/4, 26/4, 5/3, 11/7, 34/6, 40/2, 13/5, 20/6, 1/7, 30/2,
    7/7, 17/3, 24/0, 3/3, 45/3, 26/0, 30/10, 40/6, 20/2, 38/6, 36/10, 22/8}
    \draw[blue] (\x,\y) -- +(1,1);
\foreach \x/\y in {
    31/6, 37/10, 41/2, 0/7, 21/6, 27/2, 6/7, 10/7, 16/3, 35/10, 44/3,
    25/10, 2/11, 33/4, 31/2, 41/6, 20/3, 27/6, 10/3, 16/7, 14/5, 37/6,
    37/0, 21/8, 25/0, 10/9, 33/2, 6/11, 15/10, 39/10, 4/11, 2/7, 8/3,
    29/10, 41/10, 35/0, 14/3, 23/8, 36/3, 14/9, 12/9, 43/6, 25/4, 27/10,
    33/6, 12/3, 27/0, 17/10, 39/6, 21/4, 2/3, 8/7, 31/10, 29/6, 18/7,
    15/6, 21/10, 45/10, 23/4, 4/7, 10/11, 12/5, 0/3, 25/8, 45/6, 35/4,
    6/3, 19/10, 42/3, 39/2, 43/10, 21/0, 8/11, 29/2, 35/6, 33/10, 18/3,
    3/8, 1/8, 23/0, 4/3, 27/4, 33/0}
    \draw[blue] (\x,\y) +(0,1) -- +(1,0);
\end{scope}
\end{tikzpicture}}
\end{fullfigure*}

\cor{c:replace} is important because $C$ could be constructed according
to a fault tolerance threshold theorem that assumes quasi-independent error.

\begin{proof}[Proof of \thm{th:rig}] We review the definition of
quasi-independent error \cite{KLZ:resilient}.  We assume that the
edges of the circuit $C$ are subject to error.   Then we consider the
multilinear expansion \eqref{e:mpauli}.   The operator $P_J$ has a weight
$w(J)$, which is the number of tensor factors that are not the identity.
The quasi-independent error condition says that
$$\sqrt{\braket{f_J|f_J}} = O(\eps^{w(J)}).$$
for some error bound $0 < \eps < 1$.  (The fault tolerance theorem says that
fault tolerance with polylog overhead is possible if $\eps$ is small enough.
Note that Knill, Laflamme, and Zurek call this type of error \emph{monotonic}
quasi-independent error.)  

Let $P_J$ be a multi-Pauli operator that arises in the multilinear expansion
of \eqref{e:klz}.  Then we claim that
$$\sqrt{\braket{f_J|f_J}} < \delta(n)^{m(J)},$$
where $m(J)$ is the number of locally first errors in $P_J$. In each term $J$,
the state $\ket{f_J}$ is actually the entangled state $\ket{g}$ of all of 
the germs at the end of the computation.   At each position that is a 
locally first error, the norm of $\ket{g}$ decreases by a factor of 
$\delta(n)$ by \eqref{e:bound}.   At every other position, the norm does not 
increase because each operator $E_j$ is subunitary.

Finally, each edge with a locally first error has at most
$$1+k+\cdots + k^{n-1} < k^n$$
edges above it that could have errors.  It follows that
$$w(J) < k^n m(J).$$
Thus $C$ has quasi-independent error with bound
$$\eps < \delta(n)^{k^{-n}},$$
as desired.
\end{proof}

\begin{proof}[Proof of \cor{c:replace}] The original purpose of the
quasi-independent error model is that it renormalizes to itself under
any map that changes a circuit $C$ to an equivalent circuit $C'$ made
by replacing gates and qubit edges by gadgets.   This is the technique
to prove the fault tolerance theorem using concatenated quantum codes.
Such an analysis applies in our case because we can replace each edge by a
teleportation gadget, as we did in the proof of \thm{th:knill}.   Indeed,
the analysis is particularly simple because every multi-Pauli error in the
teleportation gadget in \fig{f:teleport} in the circuit $C'$ is equivalent
to a Pauli error or non-error in the original edge in $C$.   (This is a
standard fact and is left as an exercise to the reader.  Note that even
though multi-Pauli errors reduce to Pauli errors, the relative phase of
two multi-Pauli errors might change.)

Suppose that the circuit $C'$ has quasi-independent error $\eps$ by
\thm{th:main}.  Suppose that each gate in \fig{f:teleport} is available
as a single gate in the gate set.   Then 1 edge in $C$ is replaced by 8
edges in $C'$.  We suppose that $C'$ has quasi-independent error with
bound $\eps$, say by \thm{th:main}.  We suppose for simplicity that
the total error amplitude of a multi-Pauli is at most $\eps^{w(J)}$
rather than $O(\eps^{w(J)})$, although the calculation works either way.
The total amplitude of all $4^8$ multi-Pauli operators on the edges of a
teleportation gadget is at most $(1+3\eps)^8$.  One of these is the term
in which all edges are assigned $Z_0 = I$, the non-error; the other errors
are bounded by $(1+3\eps)^8 - 1$.  It follows that if we interpret $C'$
as an encoding of $C$, then $C$ has quasi-independent error with bound
$$\delta < (1+3\eps)^8 - 1.$$
(In fact, we can divide the right side by 3 by symmetry between the
non-trivial Pauli errors, but it is not necessary.)
\end{proof}

\section{A complexity class interpretation}
\label{s:complexity}

It is interesting to consider complexity classes with a bounded amount
of available quantum computation.  One example of such a class is the
\emph{Fourier hierarchy} $\FH$ \cite{Shi:tradeoff}.  In the Fourier
hierarchy, the entire circuit is quantum in the sense that it consists of
qubits, but only a bounded number of layers of Hadamard gates are allowed.
In between these layers the circuit is \emph{pseudoclassical}, meaning
that it is a unitary dilation of classical circuit.

Here we define two other mixed quantum-classical classes, even though we do
not know whether they are actually useful.  (They should not necessarily
be added to the Complexity Zoo \cite{zoo}.)  First, we can consider
the class $\SQCL$, or \emph{sandwiched quantum and classical layers}
(\fig{f:sandwich}(a)).  We represent this class by a uniform family of
polynomial-sized quantum-classical circuits $\{C_n\}$.  We assume that qubits
are only allowed in the circuit in disjoint, global layers $[t,t+b]$ that are
bounded in depth by a constant $b$.  In between the layers, all edges have
to bit edges Even though a circuit in $\SQCL$ can have a polynomial number of
layers, the fact that no quantum coherence connects any two of the layers is
a much more severe restriction than in $\FH$.  A single quantum layer is a
functional class known as $\QNC_0$, and surprisingly even this class seems
different from classical computation \cite{FGHZ:constant,BJS:collapse}.
The class $\SQCL$ can be viewed as $\BPP^{\QNC_0}$, or $\BPP$ with oracle
access to $\QNC_0$, except that it is a semantic type of oracle access in
which the oracle output is a probability distribution.

\begin{remark} We do not know whether $\QNC_0$ is weaker with noisy gates.
More precisely, whether there is a fault tolerance noise threshold below
which $\QNC_0$ is no weaker than before.
\end{remark}

We alter the definition of $\SQCL$ subtly but dramatically.  We define the
class $\AQCL$, or \emph{asynchronous quantum and classical layers} with
quantum circuits as follows:  Each qubit path in the circuit has depth
at most $b$, but the qubits do not have to disappear at the same time.
(See \fig{f:sandwich}(b).)  This definition matches the conclusion of
\thm{th:knill}, so we obtain this corollary:

\begin{corollary} $\AQCL = \BQP$.  
\end{corollary}


\begin{thebibliography}{10}

\bibitem{AB:constant}
Dorit Aharonov and Michael Ben-Or, \emph{Fault-tolerant quantum computation
  with constant error rate}, STOC '97 (El Paso, TX), ACM Press, New York, 1999,
  \mbox{arXiv:quant-ph/9906129}, pp.~176--188.

\bibitem{BBCJPW:teleport}
Charles~H. Bennett, Gilles Brassard, Claude Cr\'epeau, Richard Jozsa, Asher
  Peres, and William~K. Wootters, \emph{Teleporting an unknown quantum state
  via dual classical and einstein-podolsky-rosen channels}, Phys. Rev. Lett.
  \textbf{70} (1993), 1895--1899.

\bibitem{BJS:collapse}
Michael~J. Bremner, Richard Jozsa, and Dan~J. Shepherd, \emph{Classical
  simulation of commuting quantum computations implies collapse of the
  polynomial hierarchy}, Proc. R. Soc. Lond. Ser. A Math. Phys. Eng. Sci.
  \textbf{467} (2011), no.~2126, 459--472, \mbox{arXiv:1005.1407}.

\bibitem{FGHZ:constant}
Stephen Fenner, Frederic Green, Steven Homer, and Yong Zhang, \emph{Bounds on
  the power of constant-depth quantum circuits}, Fundamentals of computation
  theory, Lecture Notes in Comput. Sci., vol. 3623, Springer, Berlin, 2005,
  \mbox{arXiv:quant-ph/0312209}, pp.~44--55.

\bibitem{Hopper:bug}
Grace Hopper, \emph{The first bug}, Ann. Hist. Comput. \textbf{3} (1981),
  no.~3, 285--286.

\bibitem{Kalai:adversarial}
Gil Kalai, \emph{Quantum computers: noise propagation and adversarial noise
  models}, 2009, \mbox{arXiv:0904.3265}.

\bibitem{Kalai:fail}
\bysame, \emph{How quantum computers fail: quantum codes, correlations in
  physical systems, and noise accumulation}, 2011, \mbox{arXiv:1106.0485}.

\bibitem{Kitaev:imperfect}
Alexei Kitaev, \emph{Quantum error correction with imperfect gates}, Quantum
  communication, computing, and measurement (Osamu Hirota, Alexander~S. Holevo,
  and Carlton Caves, eds.), Springer, 1997, pp.~181--188.

\bibitem{Knill:postsel}
Emanuel Knill, \emph{Fault-tolerant postselected quantum computation: Schemes},
  2004, \mbox{arXiv:quant-ph/0402171}.

\bibitem{Knill:scalable}
\bysame, \emph{Scalable quantum computing in the presence of large
  detected-error rates}, Phys. Rev. A \textbf{71} (2005), 042322,
  \mbox{arXiv:quant-ph/0312190}.

\bibitem{KLZ:resilient}
Emanuel Knill, Raymond Laflamme, and Wojciech~H. Zurek, \emph{Resilient quantum
  computation: error models and thresholds}, Proc. R. Soc. Lond. A \textbf{454}
  (1998), no.~1969, 365--384, \mbox{arXiv:quant-ph/9702058}.

\bibitem{Kuperberg:memory}
Greg Kuperberg, \emph{The capacity of hybrid quantum memory}, IEEE Trans.
  Inform. Theory \textbf{49} (2003), no.~6, 1465--1473,
  \mbox{arXiv:quant-ph/0203105}.

\bibitem{NC:qcqi}
Michael~A. Nielsen and Isaac~L. Chuang, \emph{Quantum computation and quantum
  information}, Cambridge University Press, Cambridge, 2000.

\bibitem{RHG:oneway}
Robert Raussendorf, Jim Harrington, and Kovid Goyal, \emph{A fault-tolerant
  one-way quantum computer}, Ann. Physics \textbf{321} (2006), no.~9,
  2242--2270, \mbox{arXiv:quant-ph/0510135}.

\bibitem{RHG:cluster}
\bysame, \emph{Topological fault-tolerance in cluster state quantum
  computation}, New J. Phys. \textbf{9} (2007), no.~6, 199,
  \mbox{arXiv:quant-ph/0703143}.

\bibitem{Shi:tradeoff}
Yaoyun Shi, \emph{Quantum and classical tradeoffs}, Theoret. Comput. Sci.
  \textbf{344} (2005), no.~2-3, 335--345, \mbox{arXiv:quant-ph/0312213}.

\bibitem{zoo}
\emph{{The Complexity Zoo}}, \url{http://www.complexityzoo.com/}.

\end{thebibliography}

\providecommand{\bysame}{\leavevmode\hbox to3em{\hrulefill}\thinspace}
\providecommand{\MR}{\relax\ifhmode\unskip\space\fi MR }
\providecommand{\MRhref}[2]{%
  \href{http://www.ams.org/mathscinet-getitem?mr=#1}{#2}
}
\providecommand{\href}[2]{#2}

\end{document}